# Distributed GraphLab: A Framework for Machine Learning and Data Mining in the Cloud

Yucheng Low
Carnegie Mellon University
ylow@cs.cmu.edu

Joseph Gonzalez
Carnegie Mellon University
jegonzal@cs.cmu.edu

Aapo Kyrola
Carnegie Mellon University
akyrola@cs.cmu.edu

Danny Bickson
Carnegie Mellon University
bickson@cs.cmu.edu

Carlos Guestrin
Carnegie Mellon University
guestrin@cs.cmu.edu

Joseph M. Hellerstein
UC Berkeley
hellerstein@cs.berkeley.edu

## ABSTRACT

While high-level data parallel frameworks, like MapReduce, simplify the design and implementation of large-scale data processing systems, they do not naturally or efficiently support many important data mining and machine learning algorithms and can lead to inefficient learning systems. To help fill this critical void, we introduced the GraphLab abstraction which naturally expresses *asynchronous*, *dynamic*, *graph-parallel* computation while ensuring data consistency and achieving a high degree of parallel performance in the shared-memory setting. In this paper, we extend the GraphLab framework to the substantially more challenging distributed setting while preserving strong data consistency guarantees.

We develop graph based extensions to pipelined locking and data versioning to reduce network congestion and mitigate the effect of network latency. We also introduce fault tolerance to the GraphLab abstraction using the classic Chandy-Lamport snapshot algorithm and demonstrate how it can be easily implemented by exploiting the GraphLab abstraction itself. Finally, we evaluate our distributed implementation of the GraphLab abstraction on a large Amazon EC2 deployment and show 1-2 orders of magnitude performance gains over Hadoop-based implementations.

## 1. INTRODUCTION

With the exponential growth in the scale of Machine Learning and Data Mining (MLDM) problems and increasing sophistication of MLDM techniques, there is an increasing need for systems that can execute MLDM algorithms efficiently in parallel on large clusters. Simultaneously, the availability of Cloud computing services like Amazon EC2 provide the promise of on-demand access to affordable large-scale computing and storage resources without substantial upfront investments. Unfortunately, designing, implementing, and debugging the distributed MLDM algorithms needed to fully utilize the Cloud can be prohibitively challenging requiring MLDM experts to address race conditions, deadlocks, distributed state, and communication protocols while simultaneously developing mathematically complex models and algorithms.

Nonetheless, the demand for large-scale computational and storage resources, has driven many [2, 14, 15, 27, 30, 35] to develop new parallel and distributed MLDM systems targeted at individual models and applications. This time consuming and often redundant effort slows the progress of the field as different research groups repeatedly solve the same parallel/distributed computing problems. Therefore, the MLDM community needs a high-level distributed abstraction that specifically targets the *asynchronous*, *dynamic*, *graph-parallel* computation found in many MLDM applications while hiding the complexities of parallel/distributed system design. Unfortunately, existing high-level parallel abstractions (e.g. MapReduce [8, 9], Dryad [19] and Pregel [25]) fail to support these critical properties. To help fill this void we introduced [24] GraphLab abstraction which directly targets asynchronous, dynamic, graph-parallel computation in the *shared-memory* setting.

In this paper we extend the multi-core GraphLab abstraction to the distributed setting and provide a formal description of the distributed execution model. We then explore several methods to implement an efficient distributed execution model while preserving strict consistency requirements. To achieve this goal we incorporate data versioning to reduce network congestion and *pipelined distributed locking* to mitigate the effects of network latency. To address the challenges of data locality and ingress we introduce the *atom graph* for rapidly placing graph structured data in the distributed setting. We also add fault tolerance to the GraphLab framework by adapting the classic Chandy-Lamport [6] snapshot algorithm and demonstrate how it can be easily implemented within the GraphLab abstraction.

We conduct a comprehensive performance analysis of our optimized C++ implementation on the Amazon Elastic Cloud (EC2) computing service. We show that applications created using GraphLab outperform equivalent Hadoop/MapReduce[9] implementations by 20-60x and match the performance of carefully constructed MPI implementations. Our main contributions are the following:

- A summary of common properties of MLDM algorithms and the limitations of existing large-scale frameworks. (Sec. 2)
- A modified version of the GraphLab abstraction and execution model tailored to the distributed setting. (Sec. 3)
- Two substantially different approaches to implementing the new distributed execution model(Sec. 4):

- ○ **Chromatic Engine:** uses graph coloring to achieve efficient sequentially consistent execution for static schedules.
- ○ **Locking Engine:** uses pipelined distributed locking and latency hiding to support dynamically prioritized execution.
- Fault tolerance through two snapshotting schemes. (Sec. 4.3)
- Implementations of three state-of-the-art machine learning algorithms on-top of distributed GraphLab. (Sec. 5)
- An extensive evaluation of Distributed GraphLab using a 512 processor (64 node) EC2 cluster, including comparisons to Hadoop, Pregel, and MPI implementations. (Sec. 5)

## 2. MLDM ALGORITHM PROPERTIES

In this section we describe several key properties of efficient large-scale parallel MLDM systems addressed by the GraphLab abstraction [24] and how other parallel frameworks fail to address these properties. A summary of these properties and parallel frameworks can be found in Table 1.

**Graph Structured Computation:** Many of the recent advances in MLDM have focused on modeling the *dependencies* between data. By modeling data dependencies, we are able to extract more signal from noisy data. For example, modeling the dependencies between similar shoppers allows us to make better product recommendations than treating shoppers in isolation. Unfortunately, data parallel abstractions like MapReduce [9] are not generally well suited for the *dependent* computation typically required by more advanced MLDM algorithms. Although it is often possible to map algorithms with computational dependencies into the MapReduce abstraction, the resulting transformations can be challenging and may introduce substantial inefficiency.

As a consequence, there has been a recent trend toward **graph-parallel** abstractions like Pregel [25] and GraphLab [24] which naturally express computational dependencies. These abstractions adopt a vertex-centric model in which computation is defined as kernels that run on each vertex. For instance, Pregel is a bulk synchronous message passing abstraction where vertices communicate through messages. On the other hand, GraphLab is a sequential shared memory abstraction where each vertex can read and write to data on adjacent vertices and edges. The GraphLab runtime is then responsible for ensuring a consistent parallel execution. Consequently, GraphLab simplifies the design and implementation of graph-parallel algorithms by freeing the user to focus on sequential computation rather than the parallel movement of data (i.e., messaging).

**Asynchronous Iterative Computation:** Many important MLDM algorithms iteratively update a large set of parameters. Because of the underlying graph structure, parameter updates (on vertices or edges) depend (through the graph adjacency structure) on the values of other parameters. In contrast to **synchronous** systems, which update all parameters simultaneously (in parallel) using parameter values from the previous time step as input, **asynchronous** systems update parameters using the *most recent* parameter values as input. As a consequence, asynchronous systems provides many MLDM algorithms with significant algorithmic benefits. For example, linear systems (common to many MLDM algorithms) have been shown to converge faster when solved asynchronously [4]. Additionally, there are numerous other cases (e.g., belief propagation [13], expectation maximization [28], and stochastic optimization [35, 34]) where asynchronous procedures have been empirically shown to significantly outperform synchronous procedures. In Fig. 1(a) we demonstrate how asynchronous computation can substantially accelerate the convergence of PageRank.

Synchronous computation incurs costly performance penalties since the runtime of each phase is determined by the *slowest* machine. The poor performance of the slowest machine may be caused by a multitude of factors including: load and network imbalances, hardware variability, and multi-tenancy (a principal concern in the Cloud). Even in typical cluster settings, each compute node may also provide other services (e.g., distributed file systems). Imbalances in the utilization of these other services will result in substantial performance penalties if synchronous computation is used.

In addition, variability in the complexity and convergence of the individual vertex kernels can produce additional variability in execution time, even when the graph is uniformly partitioned. For example, natural graphs encountered in real-world applications have *power-law* degree distributions which can lead to highly skewed running times even with a random partition [36]. Furthermore, the actual work required for each vertex could depend on the data in a problem specific manner (e.g., local rate of convergence).

While abstractions based on bulk data processing, such as MapReduce [9] and Dryad [19] were not designed for iterative computation, recent projects such as Spark [38] extend MapReduce and other data parallel abstractions to the iterative setting. However, these abstractions still do not support asynchronous computation. Bulk Synchronous Parallel (BSP) abstractions such as Pregel [25], Piccolo [33], and BPGL [16] do not naturally express asynchronicity. On the other hand, the shared memory GraphLab abstraction was designed to efficiently and naturally express the asynchronous iterative algorithms common to advanced MLDM.

**Dynamic Computation:** In many MLDM algorithms, iterative computation converges asymmetrically. For example, in parameter optimization, often a large number of parameters will quickly converge in a few iterations, while the remaining parameters will converge slowly over many iterations [11, 10]. In Fig. 1(b) we plot the distribution of updates required to reach convergence for PageRank. Surprisingly, the majority of the vertices required only a *single* update while only about 3% of the vertices required more than 10 updates. Additionally, prioritizing computation can further accelerate convergence as demonstrated by Zhang et al. [39] for a variety of graph algorithms including PageRank. If we update all parameters equally often, we waste time recomputing parameters that have effectively converged. Conversely, by focusing early computation on more challenging parameters, we can potentially accelerate convergence. In Fig. 1(c) we empirically demonstrate how dynamic scheduling can accelerate convergence of loopy belief propagation (a popular MLDM algorithm).

Several recent abstractions have incorporated forms of dynamic computation. For example, Pregel [25] supports a limited form of dynamic computation by allowing some vertices to skip computation on each super-step. Other abstractions like Pearce et al. [32] and GraphLab allow the user to adaptively *prioritize* computation. While both Pregel and GraphLab support dynamic computation, only GraphLab permits prioritization as well as the ability to adaptively pull information from adjacent vertices (see Sec. 3.2 for more details). In this paper we relax some of the original GraphLab scheduling requirements described in [24] to enable efficient *distributed* FIFO and priority scheduling.

**Serializability:** By ensuring that all parallel executions have an equivalent sequential execution, serializability eliminates many challenges associated with designing, implementing, and testing parallel MLDM algorithms. In addition, many algorithms converge faster if serializability is ensured, and some even require serializability for correctness. For instance, Dynamic ALS (Sec. 5.1) is unstable when allowed to race (Fig. 1(d)). Gibbs sampling, a *very popular* MLDM algorithm, requires serializability for statistical correctness.

|  | Computation Model | Sparse Depend. | Async. Comp. | Iterative | Prioritized Ordering | Enforce Consistency | Distributed |
|---|---|---|---|---|---|---|---|
| MPI | Messaging | Yes | Yes | Yes | N/A | No | Yes |
| MapReduce[9] | Par. data-flow | No | No | extensions(a) | No | Yes | Yes |
| Dryad[19] | Par. data-flow | Yes | No | extensions(b) | No | Yes | Yes |
| Pregel[25]/BPGL[16] | GraphBSP | Yes | No | Yes | No | Yes | Yes |
| Piccolo[33] | Distr. map | No | No | Yes | No | Partially(c) | Yes |
| Pearce et.al.[32] | Graph Visitor | Yes | Yes | Yes | Yes | No | No |
| **GraphLab** | **GraphLab** | **Yes** | **Yes** | **Yes** | **Yes** | **Yes** | **Yes** |

**Table 1: Comparison chart of large-scale computation frameworks. (a) [38] describes and iterative extension of MapReduce. (b) [18] proposes an iterative extension for Dryad. (c) Piccolo does not provide a mechanism to ensure consistency but instead exposes a mechanism for the user to attempt to recover from simultaneous writes.**

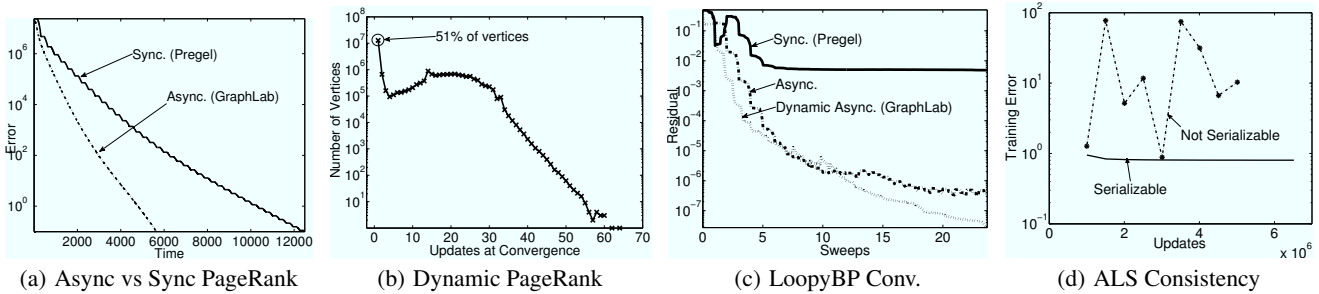

(a) Async vs Sync PageRank    (b) Dynamic PageRank    (c) LoopyBP Conv.    (d) ALS Consistency

**Figure 1: (a)** Rate of convergence, measured in $L_1$ error to the true PageRank vector versus time, of the PageRank algorithm on a 25M vertex 355M edge web graph on 16 processors. **(b)** The distribution of update counts after running dynamic PageRank to convergence. Notice that the majority of the vertices converged in only a single update. **(c)** Rate of convergence of Loopy Belief propagation on web-spam detection. **(d)** Comparing serializable and non-serializable (racing) execution of the dynamic ALS algorithm in Sec. 5.1 on the Netflix movie recommendation problem. Non-serializable execution exhibits unstable convergence behavior.

An abstraction that enforces serializable computation eliminates much of the complexity introduced by concurrency, allowing the MLDM expert to focus on the algorithm and model design. Debugging mathematical code in a concurrent program which has data-corruption caused by data races is difficult and time consuming. Surprisingly, many asynchronous abstractions like [32] do not ensure serializability or, like Piccolo [33], provide only basic mechanisms to recover from data races. GraphLab supports a broad range of consistency settings, allowing a program to choose the level of consistency needed for correctness. In Sec. 4 we describe several techniques we developed to enforce serializability in the distributed setting.

## 3. DIST. GRAPHLAB ABSTRACTION

The GraphLab abstraction consists of three main parts, the data graph, the update function, and the sync operation. The data graph (Sec. 3.1) represents user modifiable program state, and stores both the mutable user-defined data and encodes the sparse computational dependencies. The update function (Sec. 3.2) represents the user computation and operate on the data graph by transforming data in small overlapping contexts called scopes. Finally, the sync operation (Sec. 3.5) concurrently maintains global aggregates. To ground the GraphLab abstraction in a concrete problem, we will use the PageRank algorithm [31] as a running example.

EXAMPLE 1 (PAGERANK). *The PageRank algorithm recursively defines the rank of a webpage $v$:*

$$\boldsymbol{R}(v) = \frac{\alpha}{n} + (1-\alpha) \sum_{u \text{ links to } v} w_{u,v} \times \boldsymbol{R}(u) \quad (1)$$

*in terms of the weighted $w_{u,v}$ ranks $\boldsymbol{R}(u)$ of the pages $u$ that link to $v$ as well as some probability $\alpha$ of randomly jumping to that page. The PageRank algorithm iterates Eq. (1) until the PageRank changes by less than some small value $\epsilon$.*

### 3.1 Data Graph

The GraphLab abstraction stores the program state as a directed graph called the **data graph**. The data graph $G = (V, E, D)$ is a container that manages the user defined data $D$. Here we use the term *data* broadly to refer to model parameters, algorithm state, and even statistical data. Users can associate arbitrary data with each vertex $\{D_v : v \in V\}$ and edge $\{D_{u \to v} : \{u, v\} \in E\}$ in the graph. However, as the GraphLab abstraction is not dependent on edge directions, we also use $D_{u \leftrightarrow v}$ to denote the data on both edge directions $u \to v$ and $v \to u$. Finally, while the graph data is mutable, the structure is *static* and cannot be changed during execution.

EXAMPLE 2 (PAGERANK: EX. 1). *The data graph is directly obtained from the web graph, where each vertex corresponds to a web page and each edge represents a link. The vertex data $D_v$ stores $\boldsymbol{R}(v)$, the current estimate of the PageRank, and the edge data $D_{u \to v}$ stores $w_{u,v}$, the directed weight of the link.*

### 3.2 Update Functions

Computation is encoded in the GraphLab abstraction in the form of update functions. An **update function** is a stateless procedure that modifies the data within the scope of a vertex and schedules the future execution of update functions on other vertices. The **scope** of vertex $v$ (denoted by $\mathcal{S}_v$) is the data stored in $v$, as well as the data stored in all adjacent vertices and adjacent edges (Fig. 2(a)).

A GraphLab update function takes as an input a vertex $v$ and its scope $\mathcal{S}_v$ and returns the new versions of the data in the scope as well as a set vertices $\mathcal{T}$:

$$\textbf{Update} : f(v, \mathcal{S}_v) \to (\mathcal{S}_v, \mathcal{T})$$

After executing an update function the modified data in $\mathcal{S}_v$ is written back to the data graph. The set of vertices $u \in \mathcal{T}$ are *eventually* executed by applying the update function $f(u, \mathcal{S}_u)$ following the execution semantics described later in Sec. 3.3.

Rather than adopting a message passing or data flow model as in [25, 19], GraphLab allows the user defined update functions complete freedom to read and modify any of the data on adjacent vertices and edges. This simplifies user code and eliminates the need for the users to reason about the movement of data. By controlling what vertices are returned in $\mathcal{T}$ and thus to be executed, GraphLab update functions can efficiently express adaptive computation. For example, an update function may choose to return (schedule) its neighbors only when it has made a substantial change to its local data.

There is an important difference between Pregel and GraphLab in how dynamic computation is expressed. GraphLab decouples the scheduling of future computation from the movement of data. As a consequence, GraphLab update functions have access to data on adjacent vertices even if the adjacent vertices did not schedule the current update. Conversely, Pregel update functions are initiated by messages and can only access the data in the message, limiting what can be expressed. For instance, dynamic PageRank is difficult to express in Pregel since the PageRank computation for a given page requires the PageRank values of all adjacent pages even if some of the adjacent pages *have not recently changed*. Therefore, the decision to send data (PageRank values) to neighboring vertices cannot be made by the sending vertex (as required by Pregel) but instead must be made by the receiving vertex. GraphLab, naturally expresses the *pull* model, since adjacent vertices are only responsible for *scheduling*, and update functions can *directly read* adjacent vertex values even if they have not changed.

EXAMPLE 3 (PAGERANK: EX. 1). *The update function for PageRank (defined in Alg. 1) computes a weighted sum of the current ranks of neighboring vertices and assigns it as the rank of the current vertex. The algorithm is adaptive: neighbors are scheduled for update only if the value of current vertex changes by more than a predefined threshold.*

---

**Algorithm 1:** PageRank update function

**Input**: Vertex data $\mathbf{R}(v)$ from $\mathcal{S}_v$
**Input**: Edge data $\{w_{u,v} : u \in \mathbf{N}[v]\}$ from $\mathcal{S}_v$
**Input**: Neighbor vertex data $\{\mathbf{R}(u) : u \in \mathbf{N}[v]\}$ from $\mathcal{S}_v$
$\mathbf{R}_{\text{old}}(v) \leftarrow \mathbf{R}(v)$  // Save old PageRank
$\mathbf{R}(v) \leftarrow \alpha/n$
**foreach** $u \in \mathbf{N}[v]$ **do**  // Loop over neighbors
  $\quad \mathbf{R}(v) \leftarrow \mathbf{R}(v) + (1-\alpha) * w_{u,v} * \mathbf{R}(u)$
// If the PageRank changes sufficiently
**if** $|\mathbf{R}(v) - \mathbf{R}_{\text{old}}(v)| > \epsilon$ **then**
  // Schedule neighbors to be updated
  **return** $\{u : u \in \mathbf{N}[v]\}$
**Output**: Modified scope $\mathcal{S}_v$ with new $\mathbf{R}(v)$

---

## 3.3 The GraphLab Execution Model

The GraphLab execution model, presented in Alg. 2, follows a simple single loop semantics. The input to the GraphLab abstraction consists of the data graph $G = (V, E, D)$, an update function, an initial set of vertices $\mathcal{T}$ to be executed. While there are vertices remaining in $\mathcal{T}$, the algorithm removes (Line 1) and executes (Line 2) vertices, adding any new vertices back into $\mathcal{T}$ (Line 3). Duplicate vertices are ignored. The resulting data graph and global values are returned to the user on completion.

To enable a more efficient distributed execution, we relax the execution ordering requirements of the shared-memory GraphLab abstraction and allow the GraphLab run-time to determine the *best* order to execute vertices. For example, RemoveNext($\mathcal{T}$) (Line 1)

---

**Algorithm 2:** GraphLab Execution Model

**Input**: Data Graph $G = (V, E, D)$
**Input**: Initial vertex set $\mathcal{T} = \{v_1, v_2, ...\}$
**while** $\mathcal{T}$ *is not Empty* **do**
1 $\quad v \leftarrow$ RemoveNext($\mathcal{T}$)
2 $\quad (\mathcal{T}', \mathcal{S}_v) \leftarrow f(v, \mathcal{S}_v)$
3 $\quad \mathcal{T} \leftarrow \mathcal{T} \cup \mathcal{T}'$
**Output**: Modified Data Graph $G = (V, E, D')$

---

may choose to return vertices in an order that minimizes network communication or latency (see Sec. 4.2.2). The only requirement imposed by the GraphLab abstraction is that all vertices in $\mathcal{T}$ are eventually executed. Because many MLDM applications benefit from prioritization, the GraphLab abstraction allows users to assign priorities to the vertices in $\mathcal{T}$. The GraphLab run-time may use these priorities in conjunction with system level objectives to optimize the order in which the vertices are executed.

## 3.4 Ensuring Serializability

The GraphLab abstraction presents a rich *sequential model* which is automatically translated into a *parallel execution* by allowing multiple processors to execute the same loop on the same graph, removing and executing different vertices simultaneously. To retain the sequential execution semantics we must ensure that overlapping computation is not run simultaneously. We introduce several **consistency models** that allow the runtime to optimize the parallel execution while maintaining serializability.

The GraphLab runtime ensures a **serializable** execution. A serializable execution implies that there exists a corresponding serial schedule of update functions that when executed by Alg. 2 produces the same values in the data-graph. By ensuring serializability, GraphLab simplifies reasoning about highly-asynchronous dynamic computation in the distributed setting.

A simple method to achieve serializability is to ensure that the scopes of concurrently executing update functions do not overlap. In [24] we call this the **full consistency** model (see Fig. 2(b)). However, full consistency limits the potential parallelism since concurrently executing update functions must be at least two vertices apart (see Fig. 2(c)). However, for many machine learning algorithms, the update functions do not need full read/write access to all of the data within the scope. For instance, the PageRank update in Eq. (1) only requires read access to edges and neighboring vertices. To provide greater parallelism while retaining serializability, GraphLab defines the **edge consistency** model. The edge consistency model ensures each update function has exclusive read-write access to its vertex and adjacent edges but read only access to adjacent vertices Fig. 2(b)). As a consequence, the edge consistency model increases parallelism by allowing update functions with slightly overlapping scopes to safely run in parallel (see Fig. 2(c)). Finally, the **vertex consistency** model allows all update functions to be run in parallel, providing maximum parallelism.

## 3.5 Sync Operation and Global Values

In many MLDM algorithms it is necessary to maintain global statistics describing data stored in the data graph. For example, many statistical inference algorithms require tracking global convergence estimators. To address this need, the GraphLab abstraction defines global values that may be read by update functions, but are written using sync operations. Similar to aggregates in Pregel, the **sync operation** is an associative commutative sum:

$$Z = \textbf{Finalize}\left(\bigoplus_{v \in V} \textbf{Map}(\mathcal{S}_v)\right) \quad (2)$$

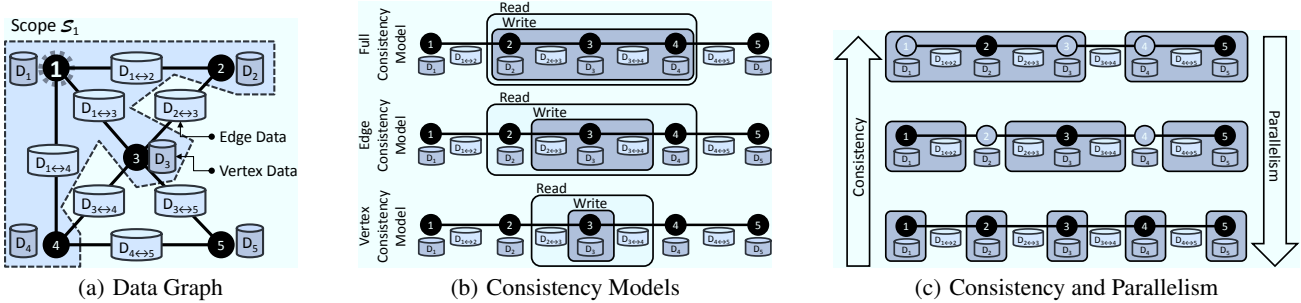

(a) Data Graph  (b) Consistency Models  (c) Consistency and Parallelism

Figure 2: (a) The data graph and scope $\mathcal{S}_1$. Gray cylinders represent the user defined vertex and edge data while the irregular region containing the vertices $\{1, 2, 3, 4\}$ is the scope, $\mathcal{S}_1$ of vertex $1$. An update function applied to vertex $1$ is able to read and modify all the data in $\mathcal{S}_1$ (vertex data $D_1$, $D_2$, $D_3$, and $D_4$ and edge data $D_{1\leftrightarrow2}$, $D_{1\leftrightarrow3}$, and $D_{1\leftrightarrow4}$). (b) The read and write permissions for an update function executed on vertex $3$ under each of the consistency models. Under the full consistency model the update function has complete read-write access to its entire scope. Under the edge consistency model, the update function has only read access to adjacent vertices. Finally, vertex consistency model only provides write access to the central vertex data. (c) The trade-off between consistency and parallelism. The dark rectangles denote the write-locked regions that cannot overlap. Update functions are executed on the dark vertices in parallel. Under the stronger consistency models fewer functions can run simultaneously.

defined over all the scopes in the graph. Unlike Pregel, the sync operation introduces a finalization phase, **Finalize**($\cdot$), to support tasks, like normalization, which are common in MLDM algorithms. Also in contrast to Pregel, where the aggregation runs after each super-step, the sync operation in GraphLab runs continuously in the background to maintain updated estimates of the global value.

Since every update function can access global values, ensuring serializability of the Sync operation with respect to update functions is costly and will generally require synchronizing and halting all computation. Just as GraphLab has multiple consistency levels for update functions, we similarly provide the option of consistent or inconsistent Sync computations.

## 4. DISTRIBUTED GRAPHLAB DESIGN

In this section we extend the shared memory system design of the GraphLab abstraction to the substantially more challenging distributed setting and discuss the techniques required to achieve this goal. An overview of the distributed design is illustrated in Fig. 5(a). Because of the inherently random memory access patterns common to dynamic asynchronous graph algorithms, we focus on the distributed **in-memory** setting, requiring the entire graph and all program state to reside in RAM. Our distributed implementation[1] is written in C++ and extends the original open-sourced shared memory GraphLab implementation.

### 4.1 The Distributed Data Graph

Efficiently implementing the data graph in the distributed setting requires balancing computation, communication, and storage. Therefore, we need to construct balanced partitioning of the data graph that minimize number of edges that cross between machines. Because the Cloud setting enables the size of the cluster to vary with budget and performance demands, we must be able to quickly load the data-graph on varying sized Cloud deployments. To resolve these challenges, we developed a graph representation based on two-phased partitioning which can be efficiently load balanced on arbitrary cluster sizes.

The data graph is initially over-partitioned using domain specific knowledge (e.g., planar embedding), or by using a distributed graph partitioning heuristic (e.g., ParMetis [21], Random Hashing) into $k$ parts where $k$ is much greater than the number of machines. Each part, called an **atom**, is stored as a separate file on a distributed storage system (e.g., HDFS, Amazon S3). Each atom file is a simple binary compressed journal of graph generating commands such as AddVertex(5000, vdata) and AddEdge(42 $\rightarrow$ 314, edata). In addition, each atom stores information regarding **ghosts**: the set of vertices and edges adjacent to the partition boundary. The connectivity structure and file locations of the $k$ atoms is stored in a **atom index** file as a **meta-graph** with $k$ vertices (corresponding to the atoms) and edges encoding the connectivity of the atoms.

[1] The open-source C++ reference implementation of the Distributed GraphLab framework is available at http://graphlab.org.

Distributed loading is accomplished by performing a fast balanced partition of the meta-graph over the number of physical machines. Each machine then constructs its local portion of the graph by playing back the journal from each of its assigned atoms. The playback procedure also instantiates the ghost of the local partition in memory. The ghosts are used as caches for their true counterparts across the network. Cache coherence is managed using a simple versioning system, eliminating the transmission of unchanged or constant data (e.g., edge weights).

The two-stage partitioning technique allows the same graph partition computation to be reused for different numbers of machines without requiring a full repartitioning step. A study on the quality of the two stage partitioning scheme is beyond the scope of this paper, though simple experiments using graphs obtained from [23] suggest that the performance is comparable to direct partitioning.

### 4.2 Distributed GraphLab Engines

The Distributed GraphLab **engine** emulates the *execution model* defined in Sec. 3.3 and is responsible for executing update functions and sync operations, maintaining the set of scheduled vertices $\mathcal{T}$, and ensuring serializability with respect to the appropriate consistency model (see Sec. 3.4). As discussed in Sec. 3.3, the precise order in which vertices are removed from $\mathcal{T}$ is up to the implementation and can affect performance and expressiveness. To evaluate this trade-off we built the low-overhead **Chromatic Engine**, which executes $\mathcal{T}$ partially asynchronously, and the more expressive **Locking Engine** which is fully asynchronous and supports vertex priorities.

#### 4.2.1 Chromatic Engine

A classic technique to achieve a serializable parallel execution of a set of dependent tasks (represented as vertices in a graph) is to construct a vertex coloring that assigns a color to each vertex

720

such that no adjacent vertices share the same color [4]. Given a vertex coloring of the data graph, we can satisfy the *edge consistency model* by executing, *synchronously*, all vertices of the same color in the vertex set $\mathcal{T}$ before proceeding to the next color. We use the term **color-step**, in analogy to the *super-step* in the BSP model, to describe the process of updating all the vertices within a single color and communicating all changes. The sync operation can then be run safely between color-steps.

We can satisfy the other consistency models simply by changing how the vertices are colored. The *full consistency model* is satisfied by constructing a second-order vertex coloring (i.e., no vertex shares the same color as any of its distance two neighbors). The *vertex consistency model* is satisfied by assigning all vertices the same color. While optimal graph coloring is NP-hard in general, a reasonable quality coloring can be constructed quickly using graph coloring heuristics (e.g., greedy coloring). Furthermore, many MLDM problems produce graphs with trivial colorings. For example, many optimization problems in MLDM are naturally expressed as bipartite (two-colorable) graphs, while problems based upon template models can be easily colored using the template [12].

While the chromatic engine operates in synchronous color-steps, changes to ghost vertices and edges are communicated asynchronously as they are made. Consequently, the chromatic engine efficiently uses both network bandwidth and processor time within each color-step. However, we must ensure that all modifications are communicated before moving to the next color and therefore we require a full communication barrier between color-steps.

### 4.2.2 Distributed Locking Engine

While the chromatic engine satisfies the distributed GraphLab abstraction defined in Sec. 3, it does not provide sufficient scheduling flexibility for many interesting applications. In addition, it presupposes the availability of a graph coloring, which may not always be readily available. To overcome these limitations, we introduce the distributed locking engine which extends the mutual exclusion technique used in the shared memory engine.

We achieve distributed mutual exclusion by associating a readers-writer lock with each vertex. The different consistency models can then be implemented using different locking protocols. Vertex consistency is achieved by acquiring a write-lock on the central vertex of each requested scope. Edge consistency is achieved by acquiring a write lock on the central vertex, and read locks on adjacent vertices. Finally, full consistency is achieved by acquiring write locks on the central vertex and all adjacent vertices. Deadlocks are avoided by acquiring locks sequentially following a canonical order. We use the ordering induced by machine ID followed by vertex ID (owner($v$), $v$) since this allows all locks on a remote machine to be requested in a single message.

Since the graph is partitioned, we restrict each machine to only run updates on local vertices. The ghost vertices/edges ensure that the update have direct memory access to all information in the scope. Each worker thread on each machine evaluates the loop described in Alg. 3 until the scheduler is empty. Termination is evaluated using the distributed consensus algorithm described in [26].

A naive implementation Alg. 3 will perform poorly due to the latency of remote lock acquisition and data synchronization. We therefore rely on several techniques to both reduce latency and hide its effects [17]. First, the ghosting system provides caching capabilities eliminating the need to transmit or wait on data that has not changed remotely. Second, all lock requests and synchronization calls are *pipelined* allowing each machine to request locks and data for many scopes simultaneously and then evaluate the update function only when the scope is ready.

**Algorithm 3:** Naive Locking Engine Thread Loop

**while** *not done* **do**
    Get next vertex $v$ from scheduler
    Acquire locks and synchronize data for scope $\mathcal{S}_v$
    Execute $(\mathcal{T}', \mathcal{S}_v) = f(v, \mathcal{S}_v)$ on scope $\mathcal{S}_v$
    // update scheduler on each machine
    For each machine $p$, Send $\{s \in \mathcal{T}' : \text{owner}(s) = p\}$
    Release locks and push changes for scope $\mathcal{S}_v$

**Algorithm 4:** Pipelined Locking Engine Thread Loop

**while** *not done* **do**
    **if** *Pipeline Has Ready Vertex $v$* **then**
        Execute $(\mathcal{T}', \mathcal{S}_v) = f(v, \mathcal{S}_V)$
        // update scheduler on each machine
        For each machine $p$, Send $\{s \in \mathcal{T}' : \text{owner}(s) = p\}$
        Release locks and push changes to $\mathcal{S}_v$ in background
    **else**
        Wait on the Pipeline

**Pipelined Locking and Prefetching:** Each machine maintains a pipeline of vertices for which locks have been requested, but have not been fulfilled. Vertices that complete lock acquisition and data synchronization leave the pipeline and are executed by worker threads. The local scheduler ensures that the pipeline is always filled to capacity. An overview of the pipelined locking engine loop is shown in Alg. 4.

To implement the pipelining system, regular readers-writer locks cannot be used since they would halt the pipeline thread on contention. We therefore implemented a non-blocking variation of the readers-writer lock that operates through callbacks. Lock acquisition requests provide a pointer to a callback, that is called once the request is fulfilled. These callbacks are chained into a distributed continuation passing scheme that passes lock requests across machines in sequence. Since lock acquisition follows the total ordering described earlier, deadlock free operation is guaranteed. To further reduce latency, synchronization of locked data is performed immediately as each machine completes its local locks.

EXAMPLE 4. *To acquire a distributed edge consistent scope on a vertex $v$ owned by machine $2$ with ghosts on machines $1$ and $5$, the system first sends a message to machine $1$ to acquire a* local *edge consistent scope on machine $1$ (write-lock on $v$, read-lock on neighbors). Once the locks are acquired, the message is passed on to machine $2$ to again acquire a* local *edge consistent scope. Finally, the message is sent to machine $5$ before returning to the owning machine to signal completion.*

To evaluate the performance of the distributed pipelining system, we constructed a three-dimensional mesh of $300 \times 300 \times 300 = 27,000,000$ vertices. Each vertex is 26-connected (to immediately adjacent vertices along the axis directions, as well as all diagonals), producing over 375 million edges. The graph is partitioned using Metis[21] into 512 atoms. We interpret the graph as a binary Markov Random Field [13] and evaluate the runtime of 10 iterations of loopy Belief Propagation [13] varying the length of the pipeline from 100 to 10,000, and the number of EC2 cluster compute instance (`cc1.4xlarge`) from 4 machines (32 processors) to 16 machines (128 processors). We observe in Fig. 3(a) that the distributed locking system provides strong, nearly linear, scalability. In Fig. 3(b) we evaluate the efficacy of the pipelining system by increasing the pipeline length. We find that increasing the length from 100 to 1000 leads to a *factor of three* reduction in runtime.

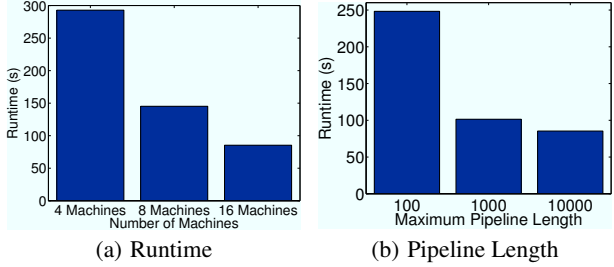

(a) Runtime  (b) Pipeline Length

**Figure 3: (a)** Plots the runtime of the Distributed Locking Engine on a synthetic loopy belief propagation problem varying the number of machines with pipeline length $= 10,000$. **(b)** Plots the runtime of the Distributed Locking Engine on the same synthetic problem on 16 machines (128 CPUs), varying the pipeline length. Increasing pipeline length improves performance with diminishing returns.

---

**Algorithm 5:** Snapshot Update on vertex $v$

**if** $v$ *was already snapshotted* **then**
  Quit
Save $D_v$ // Save current vertex
**foreach** $u \in \mathbf{N}[v]$ **do**    // Loop over neighbors
  **if** $u$ *was not snapshotted* **then**
    Save data on edge $D_{u \leftrightarrow v}$
    Schedule $u$ for a Snapshot Update
Mark $v$ as snapshotted

---

### 4.3 Fault Tolerance

We introduce fault tolerance to the distributed GraphLab framework using a distributed checkpoint mechanism. In the event of a failure, the system is recovered from the last checkpoint. We evaluate two strategies to construct distributed snapshots: a synchronous method that suspends all computation while the snapshot is constructed, and an asynchronous method that incrementally constructs a snapshot without suspending execution.

Synchronous snapshots are constructed by suspending execution of update functions, flushing all communication channels, and then saving all modified data since the last snapshot. Changes are written to journal files in a distributed file-system and can be used to restart the execution at any previous snapshot.

Unfortunately, synchronous snapshots expose the GraphLab engine to the same inefficiencies of synchronous computation (Sec. 2) that GraphLab is trying to address. Therefore we designed a fully asynchronous alternative based on the Chandy-Lamport [6] snapshot. Using the GraphLab abstraction we designed and implemented a variant of the Chandy-Lamport snapshot specifically tailored to the GraphLab data-graph and execution model. The resulting algorithm (Alg. 5) is expressed as an update function and guarantees a consistent snapshot under the following conditions:

- Edge Consistency is used on all update functions,
- Schedule completes before the scope is unlocked,
- the Snapshot Update is prioritized over other update functions,

which are satisfied with minimal changes to the GraphLab engine. The proof of correctness follows naturally from the original proof in [6] with the machines and channels replaced by vertices and edges and messages corresponding to scope modifications.

Both the synchronous and asynchronous snapshots are initiated at fixed intervals. The choice of interval must balance the cost of constructing the checkpoint with the computation lost since the last checkpoint in the event of a failure. Young et al. [37] derived a first-order approximation to the optimal checkpoint interval:

$$T_{\text{Interval}} = \sqrt{2T_{\text{checkpoint}}T_{\text{MTBF}}} \quad (3)$$

where $T_{\text{checkpoint}}$ is the time it takes to complete the checkpoint and $T_{\text{MTBF}}$ is the mean time between failures for the cluster. For instance, using a cluster of 64 machines, a per machine MTBF of 1 year, and a checkpoint time of 2 min leads to optimal checkpoint intervals of 3 hrs. Therefore, for the deployments considered in our experiments, even taking pessimistic assumptions for $T_{\text{MTBF}}$, leads to checkpoint intervals that far exceed the runtime of our experiments and in fact also exceed the Hadoop experiment runtimes. This brings into question the emphasis on strong fault tolerance in Hadoop. Better performance can be obtained by balancing fault tolerance costs against that of a job restart.

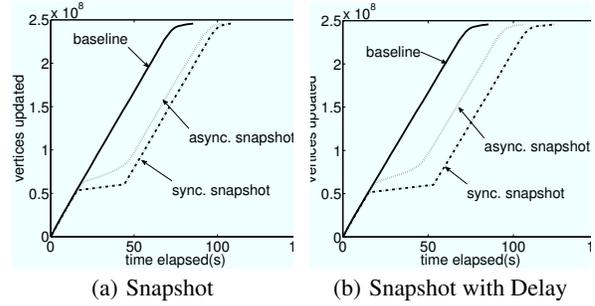

(a) Snapshot  (b) Snapshot with Delay

**Figure 4: (a)** The number of vertices updated vs. time elapsed for 10 iterations comparing asynchronous and synchronous snapshots. Synchronous snapshots (completed in 109 seconds) have the characteristic "flatline" while asynchronous snapshots (completed in 104 seconds) allow computation to proceed. **(b)** Same setup as in (a) but with a single machine fault lasting 15 seconds. As a result of the 15 second delay the asynchronous snapshot incurs only a 3 second penalty while the synchronous snapshot incurs a 16 second penalty.

**Evaluation:** We evaluate the performance of the snapshotting algorithms on the same synthetic mesh problem described in the previous section, running on 16 machines (128 processors). We configure the implementation to issue exactly one snapshot in the middle of the second iteration. In Fig. 4(a) we plot the number of updates completed against time elapsed. The effect of the synchronous snapshot and the asynchronous snapshot can be clearly observed: synchronous snapshots stops execution, while the asynchronous snapshot only slows down execution.

The benefits of asynchronous snapshots become more apparent in the **multi-tenancy** setting where variation in system performance exacerbate the cost of synchronous operations. We simulate this on Amazon EC2 by halting one of the processes for 15 seconds after snapshot begins. In figures Fig. 4(b) we again plot the number of updates completed against time elapsed and we observe that the asynchronous snapshot is minimally affected by the simulated failure (adding only 3 seconds to the runtime), while the synchronous snapshot experiences a full 15 second increase in runtime.

### 4.4 System Design

In Fig. 5(a), we provide a high-level overview of a GraphLab system. The user begins by constructing the atom graph representation on a Distributed File System (DFS). If hashed partitioning is used, the construction process is Map-Reduceable where a map is performed over each vertex and edge, and each reducer accumulates an atom file. The atom journal format allows future changes to the graph to be appended without reprocessing all the data.

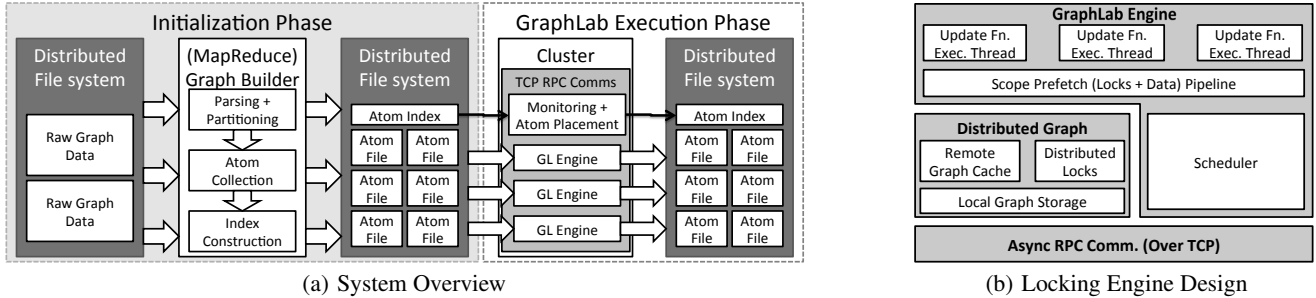

(a) System Overview                (b) Locking Engine Design

Figure 5: (a) A high level overview of the GraphLab system. In the initialization phase the atom file representation of the data graph is constructed. In the GraphLab Execution phase the atom files are assigned to individual execution engines and are then loaded from the DFS. (b) A block diagram of the parts of the Distributed GraphLab process. Each block in the diagram makes use of the blocks below it. For more details, see Sec. 4.4.

Fig. 5(b) provides a high level overview of the GraphLab locking engine implementation. When GraphLab is launched on a cluster, one instance of the GraphLab program is executed on each machine. The GraphLab processes are symmetric and directly communicate with each other using a custom asynchronous RPC protocol over TCP/IP. The first process has the additional responsibility of being a master/monitoring machine.

At launch the master process computes the placement of the atoms based on the atom index, following which all processes perform a parallel load of the atoms they were assigned. Each process is responsible for a partition of the distributed graph that is managed within a local graph storage, and provides distributed locks. A cache is used to provide access to remote graph data.

Each process also contains a scheduler that manages the vertices in $\mathcal{T}$ that have been assigned to the process. At runtime, each machine's local scheduler feeds vertices into a prefetch pipeline that collects the data and locks required to execute the vertex. Once all data and locks have been acquired, the vertex is executed by a pool of worker threads. Vertex scheduling is decentralized with each machine managing the schedule for its local vertices and forwarding scheduling requests for remote vertices. Finally, a distributed consensus algorithm [26] is used to determine when all schedulers are empty. Due to the symmetric design of the distributed runtime, there is no centralized bottleneck.

## 5. APPLICATIONS

We evaluated GraphLab on three state-of-the-art MLDM applications: collaborative filtering for Netflix movie recommendations, Video Co-segmentation (CoSeg) and Named Entity Recognition (NER). Each experiment was based on large real-world problems and datasets (see Table 2). We used the Chromatic engine for the Netflix and NER applications and the Locking Engine for the CoSeg application. Equivalent Hadoop and MPI implementations were also evaluated on the Netflix and NER applications.

Unfortunately, we could not directly compare against Pregel since it is not publicly available and current open source implementations do not scale to even the smaller problems we considered. While Pregel exposes a vertex parallel abstraction, it must still provide access to the adjacent edges within update functions. In the case of the problems considered here, the computation demands that edges be *bi-directed*, resulting in an increase in graph storage complexity (for instance, the movie "Harry Potter" connects to a very large number of users). Finally, many Pregel implementations of MLDM algorithms will require each vertex to transmit its own value to all adjacent vertices, unnecessarily expanding the amount of program state from $O(|V|)$ to $O(|E|)$.

Experiments were performed on Amazon's Elastic Computing Cloud (EC2) using up to 64 High-Performance Cluster (HPC) instances (`cc1.4xlarge`) each with dual Intel Xeon X5570 quad-core Nehalem processors and 22 GB of memory and connected by a 10 Gigabit Ethernet network. All timings include data loading and are averaged over three or more runs. On each node, GraphLab spawns eight engine threads (matching the number of cores). Numerous other threads are spawned for background communication.

In Fig. 6(a) we present an aggregate summary of the parallel speedup of GraphLab when run on 4 to 64 HPC machines on all three applications. In all cases, speedup is measured relative to the *four node deployment* since single node experiments were not always feasible due to memory limitations. No snapshots were constructed during the timing experiments since all experiments completed prior to the first snapshot under the optimal snapshot interval (3 hours) as computed in Sec. 4.3. To provide intuition regarding the snapshot cost, in Fig. 8(d) we plot for each application, the overhead of compiling a snapshot on a 64 machine cluster.

Our principal findings are:

- On equivalent tasks, GraphLab outperforms Hadoop by 20-60x and performance is comparable to tailored MPI implementations.
- GraphLab's performance scaling improves with higher computation to communication ratios.
- The GraphLab abstraction more compactly expresses the Netflix, NER and Coseg algorithms than MapReduce or MPI.

### 5.1 Netflix Movie Recommendation

The Netflix movie recommendation task uses *collaborative filtering* to predict the movie ratings for each user, based on the ratings of similar users. We implemented the **alternating least squares** (**ALS**) algorithm [40], a common algorithm in collaborative filtering. The input to ALS is a sparse users by movies matrix $R$, containing the movie ratings of each user. The algorithm iteratively computes a low-rank matrix factorization:

$$R \approx U \times V \qquad (4)$$

where $U$ and $V$ are rank $d$ matrices. The ALS algorithm alternates between computing the least-squares solution for $U$ and $V$ while holding the other fixed. Both the quality of the approximation and the computational complexity depend on the magnitude of $d$: higher $d$ produces higher accuracy while increasing computational cost. Collaborative filtering and the ALS algorithm are important tools in MLDM: an effective solution for ALS can be extended to a broad class of other applications.

While ALS may not seem like a graph algorithm, it can be represented elegantly using the GraphLab abstraction. The *sparse* matrix

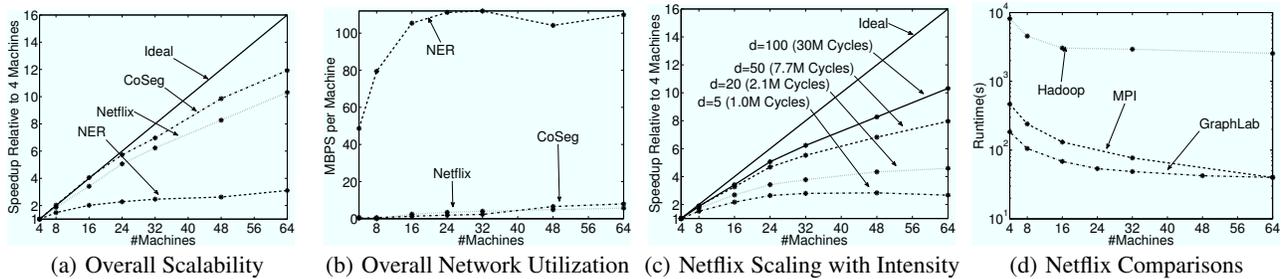

(a) Overall Scalability  (b) Overall Network Utilization  (c) Netflix Scaling with Intensity  (d) Netflix Comparisons

Figure 6: (a) Scalability of the three test applications with the largest input size. CoSeg scales excellently due to very sparse graph and high computational intensity. Netflix with default input size scales moderately while NER is hindered by high network utilization. See Sec. 5 for a detailed discussion. (b) Average bandwidth utilization per cluster node. Netflix and CoSeg have very low bandwidth requirements while NER appears to saturate when #machines $> 24$. (c) Scalability of Netflix varying the computation cost of the update function. (d) Runtime of Netflix with GraphLab, Hadoop and MPI implementations. Note the logarithmic scale. GraphLab outperforms Hadoop by 40-60x and is comparable to an MPI implementation. See Sec. 5.1 and Sec. 5.3 for a detailed discussion.

$R$ defines a bipartite graph connecting each user with the movies they rated. The edge data contains the rating for a movie-user pair. The vertex data for users and movies contains the corresponding row in $U$ and column in $V$ respectively. The GraphLab update function recomputes the $d$ length vector for each vertex by reading the $d$ length vectors on adjacent vertices and then solving a least-squares regression problem to predict the edge values. Since the graph is bipartite and two colorable, and the edge consistency model is sufficient for serializability, the chromatic engine is used.

The Netflix task provides us with an opportunity to quantify the distributed chromatic engine overhead since we are able to directly control the computation-communication ratio by manipulating $d$: the dimensionality of the approximating matrices in Eq. (4). In Fig. 6(c) we plot the speedup achieved for varying values of $d$ and the corresponding number of cycles required per update. Extrapolating to obtain the theoretically optimal runtime, we estimated the overhead of Distributed GraphLab at 64 machines (512 CPUs) to be about 12x for $d = 5$ and about 4.9x for $d = 100$. Note that this overhead includes graph loading and communication. This provides us with a measurable objective for future optimizations.

Next, we compare against a Hadoop and an MPI implementation in Fig. 6(d) ($d = 20$ for all cases), using between 4 to 64 machines. The Hadoop implementation is part of the Mahout project and is widely used. Since fault tolerance was not needed during our experiments, we reduced the Hadoop Distributed Filesystem's (HDFS) replication factor to one. A significant amount of our effort was then spent tuning the Hadoop job parameters to improve performance. However, even so we find that GraphLab performs between **40-60** times faster than Hadoop.

While some of the Hadoop inefficiency may be attributed to Java, job scheduling, and various design decisions, GraphLab also leads to a more efficient representation of the underlying algorithm. We can observe that the Map function of a Hadoop ALS implementation, performs no computation and its only purpose is to emit *copies* of the vertex data for every edge in the graph; unnecessarily multiplying the amount of data that need to be tracked.

For example, a user vertex that connects to 100 movies must emit the data on the user vertex 100 times, once for each movie. This results in the generation of a large amount of unnecessary network traffic and unnecessary HDFS writes. This weakness extends beyond the MapReduce abstraction, but also affects the graph message-passing models (such as Pregel) due to the lack of a *scatter* operation that would avoid sending same value multiple times to each machine. Comparatively, the GraphLab update function is simpler as users do not need to explicitly define the flow of information. Synchronization of a modified vertex only requires as much communication as there are ghosts of the vertex. In particular, only machines that require the vertex data for computation will receive it, and each machine receives each modified vertex data at most once, even if the vertex has many neighbors.

Our MPI implementation of ALS is highly optimized, and uses synchronous MPI collective operations for communication. The computation is broken into super-steps that alternate between recomputing the latent user and movies low rank matrices. Between super-steps the new user and movie values are scattered (using MPI_Alltoall) to the machines that need them in the next super-step. As a consequence our MPI implementation of ALS is roughly equivalent to an optimized Pregel version of ALS with added support for parallel broadcasts. Surprisingly, GraphLab was able to outperform the MPI implementation. We attribute the performance to the use of background asynchronous communication in GraphLab.

Finally, we can evaluate the effect of enabling dynamic computation. In Fig. 9(a), we plot the test error obtained over time using a dynamic update schedule as compared to a static BSP-style update schedule. This dynamic schedule is easily represented in GraphLab while it is difficult to express using Pregel messaging semantics. We observe that a dynamic schedule converges much faster, reaching a low test error in about half amount of work.

## 5.2 Video Co-segmentation (CoSeg)

Video co-segmentation automatically identifies and clusters spatio-temporal segments of video (Fig. 7(a)) that share similar texture and color characteristics. The resulting segmentation (Fig. 7(a)) can be used in scene understanding and other computer vision and robotics applications. Previous co-segmentation methods [3] have focused on processing frames in isolation. Instead, we developed a joint co-segmentation algorithm that processes all frames simultaneously and is able to model temporal stability.

We preprocessed 1,740 frames of high-resolution video by coarsening each frame to a regular grid of $120 \times 50$ rectangular **superpixels**. Each super-pixel stores the color and texture statistics for all the raw pixels in its domain. The CoSeg algorithm predicts the best label (e.g., sky, building, grass, pavement, trees) for each super pixel using **Gaussian Mixture Model** (**GMM**) in conjunction with **Loopy Belief Propagation** (**LBP**) [14]. The GMM estimates the best label given the color and texture statistics in the super-pixel. The algorithm operates by connecting neighboring pixels in time and space into a large three-dimensional grid and uses LBP to smooth the local estimates. We combined the two algorithms to form an Expectation-Maximization algorithm, alternating between LBP to compute the label for each super-pixel given the GMM and then updating the GMM given the labels from LBP.

The GraphLab update function executes the LBP local iterative update. We implement the state-of-the-art adaptive update schedule described by [11], where updates that are expected to change vertex values significantly are prioritized. We therefore make use of the locking engine with an approximate priority scheduler. The parameters for the GMM are maintained using the sync operation. To the best of our knowledge, there are no other abstractions that provide the dynamic asynchronous scheduling as well as the sync (reduction) capabilities required by this application.

In Fig. 6(a) we demonstrate that the locking engine can achieve scalability and performance on the large 10.5 million vertex graph used by this application, resulting in a 10x speedup with 16x more machines. We also observe from Fig. 8(a) that the locking engine provides nearly optimal weak scaling: the runtime does not increase significantly as the size of the graph increases proportionately with the number of machines. We can attribute this to the properties of the graph partition where the number of edges crossing machines increases linearly with the number of machines, resulting in low communication volume.

While Sec. 4.2.2 contains a limited evaluation of the pipelining system on a synthetic graph, here we further investigate the behavior of the distributed lock implementation when run on a complete problem that makes use of all key aspects of GraphLab: both sync and dynamic prioritized scheduling. The evaluation is performed on a small 32-frame (192K vertices) problem using a 4 node cluster and two different partitioning. An *optimal partition* was constructed by evenly distributing 8 frame blocks to each machine. A *worst case partition* was constructed by striping frames across machines and consequently stressing the distributed lock implementation by forcing each scope acquisition is to grab at least one remote lock. We also vary the maximum length of the pipeline. Results are plotted in Fig. 8(b). We demonstrate that increasing the length of the pipeline increases performance significantly and is able to compensate for poor partitioning, rapidly bringing down the runtime of the problem. Just as in Sec. 4.2.2, we observe diminishing returns with increasing pipeline length. While pipelining violates the priority order, rapid convergence is still achieved.

We conclude that for the video co-segmentation task, Distributed GraphLab provides excellent performance while being the only distributed graph abstraction that allows the use of dynamic prioritized scheduling. In addition, the pipelining system is an effective way to hide latency, and to some extent, a poor partitioning.

## 5.3 Named Entity Recognition (NER)

Named Entity Recognition (NER) is the task of determining the type (e.g., *Person*, *Place*, or *Thing*) of a **noun-phrase** (e.g., *Obama*, *Chicago*, or *Car*) from its **context** (e.g., "President __", "lives near __", or "bought a __"). NER is used in many natural language processing applications as well as information retrieval. In this application we obtained a large crawl of the web from the NELL project [5], and we counted the number of occurrences of each noun-phrase in each context. Starting with a small seed set of pre-labeled noun-phrases, the CoEM algorithm labels the remaining noun-phrases and contexts (see Table 7(b)) by alternating between estimating the best assignment to each noun-phrase given the types of its contexts and estimating the type of each context given the types of its noun-phrases.

The data graph for the NER problem is bipartite with one set of vertices corresponding to noun-phrases and other corresponding to each contexts. There is an edge between a noun-phrase and a context if the noun-phrase occurs in the context. The vertices store the estimated distribution over types and the edges store the number of times the noun-phrase appears in a context. Since the graph is two colorable and relatively dense the chromatic engine was used with random partitioning. The lightweight floating point arithmetic in the NER computation in conjunction with the relatively dense graph structure and random partitioning is essentially the worst-case for the current Distributed GraphLab design, and thus allow us to evaluate the overhead of the Distributed GraphLab runtime.

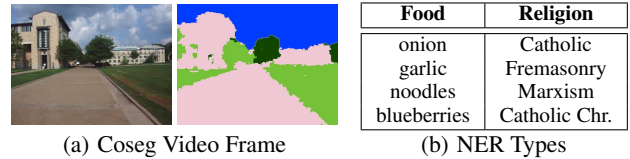

(a) Coseg Video Frame      (b) NER Types

**Figure 7: (a) Coseg: a frame from the original video sequence and the result of running the co-segmentation algorithm. (b) NER: Top words for several types.**

From Fig. 6(a) we see that NER achieved only a modest 3x improvement using 16x more machines. We attribute the poor scaling performance of NER to the large vertex data size (816 bytes), dense connectivity, and poor partitioning (random cut) that resulted in substantial communication overhead per iteration. Fig. 6(b) shows for each application, the average number of bytes per second transmitted by each machine with varying size deployments. Beyond 16 machines, NER saturates with each machine sending at a rate of over 100MB per second.

We evaluated our Distributed GraphLab implementation against a Hadoop and an MPI implementation in Fig. 8(c). In addition to the optimizations listed in Sec. 5.1, our Hadoop implementation required the use of binary marshaling methods to obtain reasonable performance (decreasing runtime by 5x from baseline).

We demonstrate that GraphLab implementation of NER was able to obtains a 20-30x speedup over Hadoop. The reason for the performance gap is the same as that for the Netflix evaluation. Since each vertex emits a copy of itself for each edge: in the extremely large CoEM graph, this corresponds to over 100 GB of HDFS writes occurring between the Map and Reduce stage.

On the other hand, our MPI implementation was able to outperform Distributed GraphLab by a healthy margin. The CoEM task requires extremely little computation in comparison to the amount of data it touches. We were able to evaluate that the NER update function requires $5.7x$ fewer cycles per byte of data accessed as compared to the Netflix problem at $d = 5$ (the hardest Netflix case evaluated). The extremely poor computation to communication ratio stresses our communication implementation, that is outperformed by MPI's efficient communication layer. Furthermore, Fig. 6(b) provides further evidence that we fail to fully saturate the network (that offers 10Gbps). Further optimizations to eliminate inefficiencies in GraphLab's communication layer should bring us up to parity with the MPI implementation.

We conclude that while Distributed GraphLab is suitable for the NER task providing an effective abstraction, further optimizations are needed to improve scalability and to bring performance closer to that of a dedicated MPI implementation.

## 5.4 EC2 Cost evaluation

To illustrate the monetary cost of using the alternative abstractions, we plot the price-runtime curve for the Netflix application in Fig. 9(b) in *log-log* scale. All costs are computed using fine-grained billing rather than the hourly billing used by Amazon EC2. The price-runtime curve demonstrates diminishing returns: the cost of attaining reduced runtimes increases faster than linearly. As a comparison, we provide the price-runtime curve for Hadoop on the same application. For the Netflix application, GraphLab is about *two orders of magnitude* more cost-effective than Hadoop.

| Exp. | #Verts | #Edges | Vertex Data | Edge Data | Update Complexity | Shape | Partition | Engine |
|------|--------|--------|-------------|-----------|-------------------|-------|-----------|--------|
| Netflix | 0.5M | 99M | $8d + 13$ | 16 | $O\left(d^3 + deg.\right)$ | bipartite | random | Chromatic |
| CoSeg | 10.5M | 31M | 392 | 80 | $O(deg.)$ | 3D grid | frames | Locking |
| NER | 2M | 200M | 816 | 4 | $O(deg.)$ | bipartite | random | Chromatic |

Table 2: *Experiment input sizes.* The vertex and edge data are measured in bytes and the $d$ in Netflix is the size of the latent dimension.

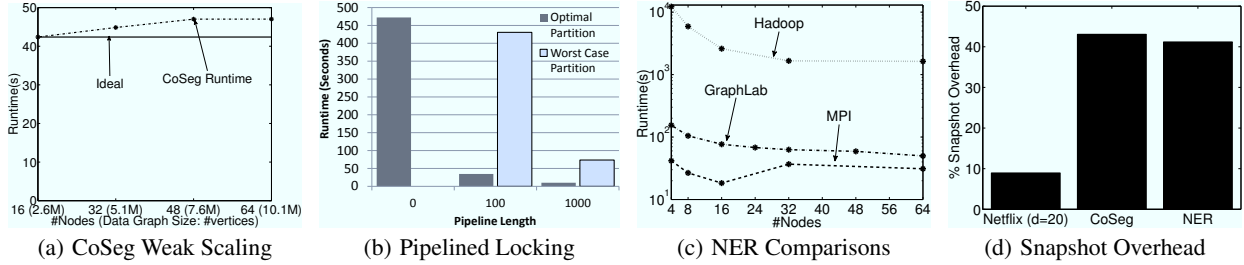

(a) CoSeg Weak Scaling  (b) Pipelined Locking  (c) NER Comparisons  (d) Snapshot Overhead

Figure 8: (a) Runtime of the CoSeg experiment as data set size is scaled proportionately with the number of machines. Ideally, runtime is constant. GraphLab experiences an 11% increase in runtime scaling from 16 to 64 machines. (b) The performance effects of varying the length of the pipeline. Increasing the pipeline length has a small effect on performance when partitioning is good. When partitioning is poor, increasing the pipeline length improves performance to be comparable to that of optimal partitioning. Runtime for worst-case partitioning at pipeline length $0$ is omitted due to excessive runtimes. (c) Runtime of the NER experiment with Distributed GraphLab, Hadoop and MPI implementations. Note the logarithmic scale. GraphLab outperforms Hadoop by about 80x when the number of machines is small, and about 30x when the number of machines is large. The performance of Distributed GraphLab is comparable to the MPI implementation. (d) For each application, the overhead of performing a complete snapshot of the graph every $|V|$ updates (where $|V|$ is the number of vertices in the graph), when running on a 64 machine cluster.

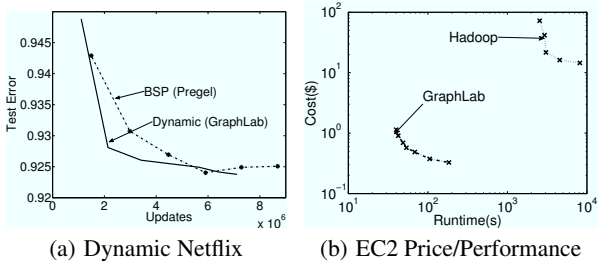

(a) Dynamic Netflix  (b) EC2 Price/Performance

Figure 9: (a) Convergence rate when dynamic computation is used. Dynamic computation can converge to equivalent test error in about half the number of updates. (b) Price Performance ratio of GraphLab and Hadoop on Amazon EC2 HPC machine on a log-log scale. Costs assume fine-grained billing.

## 6. RELATED WORK

Section 2 provides a detailed comparison of several contemporary high-level parallel and distributed frameworks. In this section we review related work in classic parallel abstractions, graph databases, and domain specific languages.

There has been substantial work [1] in graph structured databases dating back to the 1980's along with many recent open-source and commercial products (e.g., Neo4J [29]). Graph databases typically focus on efficient storage and retrieval of graph structured data with support for basic graph computation. In contrast, GraphLab focuses on iterative graph structured computation.

There are several notable projects focused on using MapReduce for graph computation. Pegasus [20] is a collection of algorithms for mining large graphs using Hadoop. Surfer [7] extends MapReduce with a *propagation* primitive, but does not support asynchronous or dynamic scheduling. Alternatively, large graphs may be "filtered" (possibly using MapReduce) to a size which can be processed on a single machine [22]. While [22] was able to derive reductions for some graph problems (e.g., minimum spanning tree), the techniques are not easily generalizable and may not be applicable to many MLDM algorithms.

## 7. CONCLUSION AND FUTURE WORK

Recent progress in MLDM research has emphasized the importance of sparse computational dependencies, asynchronous computation, dynamic scheduling and serializability in large scale MLDM problems. We described how recent distributed abstractions fail to support all three critical properties. To address these properties we introduced Distributed GraphLab, a graph-parallel distributed framework that targets these important properties of MLDM applications. Distributed GraphLab extends the shared memory GraphLab abstraction to the distributed setting by refining the execution model, relaxing the scheduling requirements, and introducing a new distributed data-graph, execution engines, and fault-tolerance systems.

We designed a **distributed data graph** format built around a two-stage partitioning scheme which allows for efficient load balancing and distributed ingress on variable-sized cluster deployments. We designed two GraphLab engines: a **chromatic engine** that is partially synchronous and assumes the existence of a graph coloring, and a **locking engine** that is fully asynchronous, supports general graph structures, and relies upon a novel **graph-based pipelined locking** system to hide network latency. Finally, we introduced two **fault tolerance** mechanisms: a synchronous snapshot algorithm and a fully asynchronous snapshot algorithm based on Chandy-Lamport snapshots that can be expressed using regular GraphLab primitives.

We implemented distributed GraphLab in C++ and evaluated it on three state-of-the-art MLDM algorithms using real data. The evaluation was performed on Amazon EC2 using up to 512 processors in 64 HPC machines. We demonstrated that Distributed GraphLab significantly outperforms Hadoop by 20-60x, and is competitive with tailored MPI implementations. We compared against BSP (Pregel) implementation of PageRank, LoopyBP, and ALS and demonstrated how support for dynamic asynchronous computation can lead to substantially improved convergence.

Future work includes extending the abstraction and runtime to support dynamically evolving graphs and external storage in graph databases. These features will enable Distributed GraphLab to continually store and processes the time evolving data commonly found in many real-world applications (e.g., social-networking and recom-mender systems). Finally, we believe that dynamic asynchronous graph-parallel computation will be a key component in large-scale machine learning and data-mining systems, and thus further research into the theory and application of these techniques will help define the emerging field of *big learning*.

*Acknowledgments*

This work is supported by the ONR Young Investigator Program grant N00014-08-1-0752, the ARO under MURI W911NF0810242, the ONR PECASE-N00014-10-1-0672, the National Science Foundation grant IIS-0803333 as well as the Intel Science and Technology Center for Cloud Computing. Joseph Gonzalez is supported by a Graduate Research Fellowship from the National Science Foundation and a fellowship from AT&T Labs.